# Distinguishing Surface and Bulk Electromagnetism via Their Dynamics in an Intrinsic Magnetic Topological Insulator

Khanh Duy Nguyen,[1] Woojoo Lee,[1,†] Jianchen Dang,[2] Tongyao Wu,[2] Gabriele Berruto,[1] Chenhui Yan,[1] Chi Ian Jess Ip,[1,‡] Haoran Lin,[1] Qiang Gao,[1] Seng Huat Lee,[3] Binghai Yan,[4] Chaoxing Liu,[3] Zhiqiang Mao,[3] Xiao-Xiao Zhang,[2] Shuolong Yang[1]*

[1]Pritzker School of Molecular Engineering, The University of Chicago, Chicago, IL 60637, USA.
[2]Department of Physics, University of Florida, Gainesville, FL 32611, USA.
[3]Department of Physics, Pennsylvania State University, University Park, PA 16802, USA.
[4]Department of Condensed Matter Physics, Weizmann Institute of Science, Rehovot 7610001, Israel.
[†]Current affiliation: Quantum Technology Institute, Korean Research Institute of Standards and Science, Daejeon 34113, Republic of Korea.
[‡]Current affiliation: Department of Physics, Massachusetts Institute of Technology, Cambridge, MA 02139, USA.
*Corresponding author: yangsl@uchicago.edu

**Abstract**

The indirect exchange interaction between local magnetic moments via surface electrons has been long predicted to bolster the surface ferromagnetism in magnetic topological insulators (MTIs), which facilitates the quantum anomalous Hall effect. This unconventional effect is critical to determining the operating temperatures of future topotronic devices. However, the experimental confirmation of this mechanism remains elusive, especially in intrinsic MTIs. Here we combine time-resolved photoemission spectroscopy with time-resolved magneto-optical Kerr effect measurements to elucidate the unique electromagnetism at the surface of an intrinsic MTI $MnBi_2Te_4$. Theoretical modeling based on 2D Ruderman-Kittel-Kasuya-Yosida interactions captures the initial quenching of a surface-rooted exchange gap within a factor of two but over-estimates the bulk demagnetization by one order of magnitude. This mechanism directly explains the sizable gap in the quasi-2D electronic state and the nonzero residual magnetization in even-layer $MnBi_2Te_4$. Furthermore, it leads to efficient light-induced demagnetization comparable to state-of-the-art magnetophotonic crystals, promising an effective manipulation of magnetism and topological orders for future topotronics.

**Introduction**

Bringing magnetism to the itinerant electronic states on the surface of 3D topological insulators (TIs) is foundational to a variety of low-dimensional topological orders such as the quantum anomalous Hall insulators (QAHI) (*1, 2*) and axion insulators (*3–5*). The magnetism in 3D TIs can be established via Anderson-Goodenough-Kanamori superexchange (*6, 7*), valence electrons (Van Vleck paramagnetism) (*8, 9*), or magnetic proximity coupling (*10*). However, the unconventional Ruderman-Kittel-Kasuya-Yosida (RKKY) interaction on the material surface is exclusively required for the time-reversal ($T$) symmetry breaking on the topological surface states (TSSs) in magnetic TIs (MTIs) (*2, 10–14*). This mechanism has been predicted to enhance the surface ferromagnetism of 3D MTIs, where the itinerant Dirac fermions with vanishing Fermi momenta strongly favor ferromagnetic coupling between magnetic moments (*11–17*). The effect is further boosted when the magnetic moments are densely and uniformly distributed as in intrinsic MTIs. Thus, the 2D RKKY interaction fundamentally determines the size of the $T$-symmetry-broken energy gap, and consequently the temperature scale at which the low-dimensional topological orders can operate. A quantitative experimental revelation of the 2D RKKY interaction on the surface of intrinsic MTIs is of fundamental importance to the study of low-dimensional topological orders, and to the ultimate material engineering for applications at realistic temperatures.

Even though there have been discussions of the RKKY interactions in several magnetically doped TI systems (*18–20*), the surface 2D RKKY interaction has not been observed directly and exclusively in intrinsic MTIs where the QAHE is expected to be realized at higher temperatures. Revealing this interaction in MTIs can be a substantial challenge using equilibrium spectroscopies, as magnetic interactions of various origins can all contribute to the overall magnetism (*21*). Here, we combine time- and angle-resolved photoemission spectroscopy (trARPES) and time-resolved magneto-optical Kerr effect (trMOKE) to reveal this distinct mechanism contributing to the surface magnetism in $MnBi_2Te_4$ (MBT), a platform on which the QAHE has been realized (*2*): a quasi-two-dimensional state ($q$-2DS) mediates the surface 2D RKKY interaction via $p$-$d$ coupling on the top MBT layer. While trARPES resolves the dynamics of the exchange gap in the $q$-2DS with meV-scale precisions, trMOKE observes the evolution of the magnetization with a dominant contribution from the bulk. Both quantities undergo a rapid quenching within 500 fs, suggesting the electronic nature of the demagnetization process. Layer-encoded frequency-domain ARPES on related $MnBi_{2n}Te_{3n+1}$ compounds (*22*) allows us to identify the surface nature of the $q$-2DS. We construct a 2D RKKY model involving localized Mn $3d$ moments and itinerant $p$ electrons. The 2D RKKY framework not only accounts for the rapid quenching of the magnetization and the exchange gap, but also provides a direct explanation for the considerably large exchange gap in the $q$-2DS. Furthermore, it can reconcile several open problems in intrinsic MTIs represented by MBT. These include the anomalously small gap at the Dirac point of the TSSs (*23–27*) and the nonzero residual magnetization in even-layer MBT (*28, 29*). Our work highlights the special magnetic interactions in the surface layer of MBT and establishes the physics foundation for effective ultrafast manipulation of magnetism in tandem with topological orders through the *p-d* interactions.

## Results

MBT hosts A-type antiferromagnetism (AFM) with the Mn 3$d$ moments ordered ferromagnetically within each septuple layer (SL), yet antiferromagnetically across adjacent SLs (Fig. 1**A**). The dynamics of the electronic band structure upon optical excitation is shown in Fig. 1. The static ARPES spectrum in Fig. 1**C** displays the typical band structure of MBT. The Dirac point of the TSS remains gapless at all temperatures within our energy resolution. The second derivative plot in Fig. 1**D** clearly shows the splitting of the $q$-2DS near -0.2 eV (*23, 26*). These two bands merge into one once the temperature is elevated above the Néel temperature $T_N$ = 25 K, as shown in Fig. 1**E**. The $q$-2DS band splitting is thus attributed to the magnetic exchange interaction. The right panel of Fig. 1**B** displays the energy distribution curve (EDC) taken at $\bar{\Gamma}$, with the $S_1$ and $S_2$ peaks further illustrating the splitting of the $q$-2DS. The second derivative plot (Fig. 1**D**) also reveals a Rashba band splitting with the band bottom near -0.1 eV (*24*). These observations generally agree with the previous studies on MBT (*23, 24*). In this report, to study the interaction between the electronic and magnetic degrees of freedom, we focus on the evolution of the exchange gap of the $q$-2DS under optical excitation. Fig. 1**F** displays the temporal evolution of the band structure upon 1.5 eV ultrafast optical excitation with an incident fluence of 10 μJ/cm$^2$ at the base temperature of 12 K. Because of the compromised energy resolution in trARPES (Methods) and of the transient spectral broadening, the $S_1$-$S_2$ band splitting is less pronounced in the trARPES spectra. Meanwhile, second-derivative plots in Fig. 1**G** suggest the existence of band splitting at all delays. This is in stark contrast to the experimental results obtained at the base temperature of 42 K (Fig. 1**H**), where a single $q$-2DS is always identified. The comparison between Fig. 1**G** and 1**H** illustrates that optical excitation at 12 K broadens the spectral features corresponding to the split $q$-2DS, rather than completely destroying the magnetic exchange gap. On the other hand, the strong diffuse photoemission intensities prevent reliable quantification of an exchange gap in the Rashba states. However, its existence is suggested by the second-derivative maps of the trARPES spectra at 360 fs when the intensity curvature near $E_F$ is suppressed. The dispersions revealed in the second-derivative map at 12 K (Fig. 1**G**) hint that a gap may open at the crossing points of the two split Rashba bands near -0.05 eV, compared to the data at 42 K (Fig. 1**H**).

To quantify the exchange gap at each time delay, we present a detailed analysis of the time-dependent EDCs taken at $k$ = 0 in Fig. 2. We first focus on the EDCs taken at 12 K (solid balls in Fig. 2**A**). The EDC taken from the static ARPES spectrum exhibits two clear peaks near -0.23 and -0.19 eV, which correspond to the $S_1$ and $S_2$ bands marked in Fig. 1**B**. Similarly, time-dependent EDCs taken at -620, 1380, and 5400 fs also exhibit two bumps corresponding to these two bands. We note that at 7 and 360 fs, this two-bump feature may not be clearly identified. However, these EDCs are distinct from the counterparts taken at 42 K (dashed lines) which exhibit a clear single-peak feature near -0.22 eV. It appears that the position of this single $q$-2DS shifts to a slightly higher binding energy instead of residing in the middle between $S_1$ and $S_2$. This is due a slight doping change in the sample used for the higher-temperature measurements. Fitting the EDCs at 12 K to a five-Lorentzian model yields time-dependent exchange gaps shown in Fig. 2**C**. Notably, for pump fluences of 10 and 20 μJ/cm$^2$ the exchange gaps are transiently reduced by ~5% and ~8%, respectively. This result shows that the functional shapes of the EDCs at 7 and 360 fs in the energy range of -0.17 ~ -0.25 eV are best understood as two broadened peaks corresponding to the $S_1$ and $S_2$ bands. Furthermore, since the exchange splitting gap between $S_1$ and $S_2$ is a manifestation of the magnetic ordering in MBT, our results give spectroscopic evidence that the magnetic subsystem is mildly affected using an infrared (IR) pump fluence up to 20 μJ/cm². We notice that the gap values obtained in the trARPES experiment using an IR fluence of 20 μJ/cm² are consistently smaller than the corresponding ones using 10 μJ/cm² (Fig. 2**C**). This is due to the steady-state heating by the IR laser resulting in an elevated sample temperature before time zero, which we estimate to be ~16 K for 10 μJ/cm² and ~21 K for 20 μJ/cm² (Supplementary Text 1). In contrast to previous trARPES studies (*24*, *30*), our ultrahigh quality trARPES experiment resolves the meV-scale dynamical change of this magnetic exchange gap. This is due to a combination of ultrahigh crystal qualities, meticulously optimized time and energy resolutions, and a high signal-to-noise ratio enabling the detailed fitting analysis.

To compare the dynamics of the magnetic and electronic subsystems, we extract the transient electronic temperature ($T_e$) using trARPES and compare it with the magnetization measured by trMOKE. By integrating EDCs over the range of [-0.2, 0.2] Å⁻¹ along $\overline{\Gamma} - \overline{K}$, we obtain the overall electron population distribution of MBT near the Fermi level ($E_F$). After the initial electron-electron thermalization in the first 300 fs, the population distribution can be described by a modified Fermi-Dirac (FD) function, which allows us to extract $T_e$ (fig. S1). Fig. 3**B** summarizes the transient $T_e$ for pump fluences of 10 and 20 μJ/cm². It reaches the peak values of 550 K for 20 μJ/cm² and 270 K for 10 μJ/cm², which are much higher than $T_N$ ~ 25 K. Moreover, $T_e$ relaxes to near the equilibrium value within ~2 ps, in contrast to a previous ultrafast electron diffuse scattering measurement reporting a substantially elevated $T_e$ even after 10 ps (*31*). This difference can be understood by considering that the measurement in Ref. (*31*) used bulk-sensitive, 3.7 MeV electron beams, whereas our ARPES measurements use 6 eV photons and are relatively surface sensitive. The ultrafast dynamics of the magnetization is tracked by trMOKE (Fig. 3**E**). Notably, even though trMOKE using 1.5 eV probe predominantly reflects the magnetic dynamics in the bulk, the initial demagnetization time scale of ~500 fs matches the onset time scale of $T_e$ as resolved by surface-sensitive trARPES, indicating a connection between the conduction band electrons and magnetic interactions. We note that the higher pump fluences in trMOKE do not cause substantial lattice heating due to the much smaller laser illumination area with the focused ~ 1.5 μm spot, which assists efficient thermal relaxation into colder regions. This is also verified by the temperature dependence of reflective magnetic circular dichroism (RMCD) when using the pulsed laser as the excitation source, with similar fluences and pulse conditions as used in trMOKE, that shows the expected magnetic transition (fig. S9). While the direct 3*d* electronic transition occurs at a higher energy compared to the 1.5 eV probe light, optical spectroscopies using < 2 eV photons still reflect the Mn magnetic moments (*29*, *32*). In particular, the optical Kerr signal has contributions from both the real and imaginary parts of the dielectric function. The real part contains a broad response from excitations at all frequencies, and, therefore the higher-energy 3*d* transitions (*33*, *34*). The applied external magnetic field (1 T) in our trMOKE measurements does not change the AFM ground state of MBT (*29*, *35*), and thus the revealed demagnetization dynamics is intrinsic to the material.

We clarify the origin of the *q*-2DS by looking into existing studies in the literature. First, static ARPES studies suggested that the *q*-2DS in MBT evolves into a pair of Rashba-split states in MnBi$_{2n}$Te$_{3n+1}$ superlattices, with the binding energy systematically increased as a function of the superlattice order *n* (*22*, 23). Second, our previous study using layer-encoded, frequency-domain ARPES on MnBi$_4$Te$_7$ elucidated that these *q*-2DSs are mostly spatially confined to the top MBT layer, based on the selective coupling between the *q*-2DS and the MBT-derived $A_{1g}$ phonon (*22*). Notably, the time scale of the $T_e$ relaxation of the *q*-2DS in MnBi$_2$Te$_4$/(Bi$_2$Te$_3$)$_n$ thin films is similar to that in MBT single crystals, but substantially different from that in Bi$_2$Te$_3$ films (Supplementary Text 3). This observation further corroborates the literature on the fact that the *q*-2DS of MBT single crystals is predominantly localized in the top MBT layer. This layer origin implies that the *q*-2DS is only sensitive to ferromagnetic ordering in the top layer, which reconciles the sizable exchange gap as seen in Figs. 1 and 2.

The insight of the spatial location of the *q*-2DS renders the top layer of MBT a unique system for magnetic interactions: the *q*-2DS effectively mediates the indirect exchange interaction between the Mn 3*d* local moments in a 2D layer, which is, by definition, the 2D RKKY interaction (*36–38*). In addition, previous ARPES studies (*24*, *39*) have resolved a 30 meV exchange gap in the Rashba-split states at the binding energy of nearly 0.1 eV, suggesting that these states also contribute to the RKKY interactions. One should note that other interaction channels such as Anderson-Goodenough-Kanamori superexchange via *p-d* orbital mixing (*6*, *7*) and the Van Vleck mechanism via the valence electrons (*8*, *9*) must not be neglected for the full picture of magnetism in MBT. In our study, we focus on the ultrafast processes driven by the excitation of conduction electrons, leading to a transient change of the 2D RKKY interaction.

To elucidate the role of RKKY physics in this system, we set up a 2D RKKY model for the topmost SL of MBT, as illustrated in Fig. 3**A**. Taking into account the available ARPES data, we assume a nearly free-electron-like conduction band of Bi-*p* and Te-*p* characters, while fixing the dispersion-less *d*-electrons to the binding energy of 4 eV (*40*). This is a simplification of the MBT system, as both the *q*-2DS and the Rashba-split states contribute to the RKKY interaction. In this model the effective magnetic Hamiltonian is:

$$H_{\text{RKKY}} = \sum_{i \neq j} \Im_{ij} \mathbf{S}_i \cdot \mathbf{S}_j \tag{1}$$

Here, the RKKY coupling constant $\Im_{ij}$ between the magnetic moments $\mathbf{S}_i$ and $\mathbf{S}_j$ at the lattice sites $\mathbf{r}_i$ and $\mathbf{r}_j$ respectively, is formulated as follows (*41–43*):

$$\Im_{ij} = -\left(\frac{J(T)}{N}\right)^2 \sum_{\mathbf{k},\mathbf{q}} \cos[(\mathbf{k}-\mathbf{q}) \cdot (\mathbf{r}_i - \mathbf{r}_j)] \frac{n_\mathbf{k} - n_\mathbf{q}}{\varepsilon_\mathbf{q} - \varepsilon_\mathbf{k}}, \tag{2}$$

where, *N* is the number of lattice sites; $\mathbf{k}$ and $\mathbf{q}$ are the momenta of *p* electrons; function *n* represents the occupation number of an electronic state where $\varepsilon$ is the corresponding binding energy. We let $n_\mathbf{k}$ follow the FD statistics with a double degeneracy, which contributes to the temperature dependence of the formalism. The Kondo coupling constant *J* defines the coupling strength $J_{pd}$ between the *p* and *d* electrons (*44*), which directly corresponds to the exchange gap in the conduction band observed by ARPES. As MBT is a cooperative magnetic system, the formation of Kondo singlets is excluded (*45*). Therefore, it is possible to use Anderson's Poor-Man scaling approach (*43–46*) to track the evolution of this *p-d* coupling strength $J_{pd}(T)$ with respect to the electronic temperature:

$$J_{pd}(T) \equiv \rho J(T) \approx \rho J + 2(\rho J)^2 \ln\left(\frac{D}{T}\right) + \mathcal{O}[(\rho J)^3], \tag{3}$$

in which, $\rho$ is the density of states of the conduction band and $D$ is the corresponding half bandwidth (Supplementary Text 4). Finally, the total temperature dependency of the RKKY coupling strength becomes

$$J_{\text{RKKY}}(T) \propto \sum_{i \neq j} -\left(\frac{J_{pd}(T)}{N}\right)^2 \sum_{k,q} \cos[(\boldsymbol{k}-\boldsymbol{q})\cdot(\boldsymbol{r}_i-\boldsymbol{r}_j)] \frac{n_k - n_q}{\varepsilon_q - \varepsilon_k}. \tag{4}$$

The calculated results show that the temporal evolution of $J_{pd}$ and $J_{\text{RKKY}}$ closely follow that of the transient electronic temperature (Fig. 3**C**). The negative values of $J_{\text{RKKY}}$ support the ferromagnetic ground state of MBT in a single SL (Supplementary Text 4). We note that the RKKY interaction can also be viewed as the weak-coupling limit of the generic *p-d* exchange mechanism of ferromagnetism in dilute magnetic semiconductors (*47*).

A direct comparison of the temporal evolutions of $J_{pd}$ and $J_{\text{RKKY}}$ with that of the exchange gap (Fig. 3**D**) and that of magnetization (Fig. 3**E**) elucidates the RKKY physics right after time zero. Indeed, all four quantities reach their minima within the first 500-1000 fs. This fast response resembles that in 2D metallic ferromagnets Fe$_x$GeTe$_2$ ($x = 3$-$5$) (*48–50*) and differs from the slower demagnetization normally seen in 2D ferromagnetic insulators at the low fluence limit, where RKKY coupling is generally unavailable (*51*). Theoretically, $J_{pd}$ is proportional to the exchange gap (*52*). The 16% maximum reduction of the calculated $J_{pd}$ is twice the maximum reduction of the exchange gap observed by trARPES (~8%). However, in this model, we use the overall spectral width of the Mn 3*d* band to estimate the bare resonant width $\Delta$ (Supplementary Text 4). This is likely an overestimation of $\Delta$ due to extrinsic contributions to the spectral width such as electron-impurity scattering, which may subsequently lead to an exaggeration of the transient reduction of $J_{pd}$. Considering these complications and the large uncertainties in the observed gap dynamics, we conclude that the 2D RKKY model sufficiently describes the initial quenching of the exchange gap in the *q*-2DS. Meanwhile, $J_{\text{RKKY}}$ is theoretically connected to the magnetization measured by trMOKE, but the connection is less straightforward. The calculated $J_{\text{RKKY}}$ is reduced by 30% under the IR fluence of 20 μJ/cm$^2$. Assuming that the 2D RKKY interaction is the only magnetic interaction in the Weiss model for ferromagnetism, this leads to ~10% reduction in the magnetization (Supplementary Text 5). However, using the linear fluence dependence of the demagnetization magnitude in trMOKE (fig. S8), and considering the different pumping and probing depths in trARPES and trMOKE, we obtain ~1.9% demagnetization in the trMOKE experiment for the same absorbed energy density as used in trARPES (Supplementary Text 5). Even after this proper normalization, the demagnetization magnitude from trMOKE is still one order of magnitude smaller than the theoretical prediction, which can be understood as follows. The magnetic dynamics probed by trMOKE using 1.5 eV light reflects magnetic interactions predominantly in the bulk. The substantial difference between the theoretical and experimental demagnetization percentages suggests that RKKY interactions make a much smaller contribution to the bulk magnetic order of MBT, as compared to the surface of MBT which hosts the *q*-2DS to mediate the surface RKKY coupling. Instead, superexchange and Van Vleck mechanisms can be the more dominant channels in the bulk, and are mostly $T_e$ independent in sub-picosecond dynamics (Supplementary Text 5) (*8, 53*). At later time (> 1 ps) the calculated coupling constants quickly return to the equilibrium values following the $T_e$ dynamics, yet the experimental exchange gap and magnetization exhibit prolonged relaxation dynamics (fig. S11). This reflects the delayed lattice heating and its impact on the orbital overlap in the superexchange interaction (*53*).

## Discussion

We have used a 2D RKKY model to successfully account for the initial demagnetization time scale as well as the order-of-magnitude for the exchange gap quenching. Notably, the combination of trARPES and trMOKE experiments confines the theoretical picture. We have also considered the Elliott-Yafet-type spin-flip model, which is widely used for metallic ferromagnets (*54*). Using our experimental transient $T_e$ and material parameters in the literature (*31*, *54*, *55*) as inputs, this model would lead to 100% demagnetization which is inconsistent with either our trARPES or trMOKE results (Supplementary Text 6). Another theoretical framework widely adopted by studies on 2D magnets is the Landau-Lifshitz-Gilbert model (*31*, *49*, *51*), in which the leading terms can be equivalent to the RKKY Hamiltonian if explicitly incorporating the *p-d* interactions (Supplementary Text 7) (*31*). The success of the 2D RKKY model in explaining the time scale and magnitude of the exchange gap reduction also corroborates the fact that the *q*-2DS resides predominantly on the top layer of MBT. The formation of the *q*-2DS can be fundamentally driven by surface defects (*56*). The same defects can lead to the relocation of the TSS into the space between the first and second SLs (*57*). The opposite magnetic moments from the first and second SLs may give rise to a vanishing *T*-symmetry-broken gap on the TSS. This, in fact, hints to a direction for future designs of high-$T_c$ QAHE, in which the TSSs should be spatially confined to the top SL by mitigating the surface defects via chemical or thermal treatments to enhance the interaction with the magnetism.

Moreover, the additional RKKY interactions mediated by the *q*-2DS suggest that the surface SL is magnetically inequivalent to the interior SLs. In ultrathin MBT flakes, the electronic structure at the top vacuum/MBT interface is expected to be different from that at the bottom MBT/substrate interface, giving rise to disparate magnetic interactions on the top and bottom surfaces. This may lead to an inversion symmetry breaking and thus provide an explanation for the residual magnetization at zero field and the hysteresis loop in even-layer MBT, characterized by RMCD (*29*, *32*) and anomalous Hall effect measurements (*28*, *58*). As the external field is scanned toward 0 T, the surface magnetization boosted by the 2D RKKY interactions on the top surface may not be compensated by that on the bottom surface, leading to an overall residual magnetization.

The 2D RKKY mechanism suggests a unique path for effective magnetic manipulation through the surface electrons, which can be relevant for future device applications. First, 2D electronic systems generally have weak dielectric screening, leading to enhanced interactions between electromagnetic fields and individual charge carriers (*59*). This aspect, together with the miniscule electronic heat capacity due to the small Fermi surfaces, gives rise to an effective electronic heating to a peak temperature > 500 K even using a mild fluence of 20 μJ/cm². Furthermore, in MBT the absorbed energy is rapidly transferred to the magnetic system through the *d-p* interaction (see Fig. 4 for a comparison with other magnets). The efficiency of modulating the magnetism in MBT, characterized by the percentage of magnetization changes normalized by the incident fluence, is higher than those of many common magnets (*60–71*) and comparable to that of engineered magnetophotonic crystals (*66*). This accelerated channel of energy transfer from optical excitation to magnetic subsystems in MBT is particularly beneficial in applications such as optoelectronics and magnetic memories toward the 2D limit. Moreover, MBT is a magnetic topological insulator where the magnetism and topology of the electronic structure are mutually locked. Recently, the axion optical induction of antiferromagnetic domains has been demonstrated in even-layer MBT flakes (*72*). Here, the revealed 2D RKKY interaction can be involved in this process and can potentially facilitate a more energy-efficient optoelectronic switching of both even-layer and odd-layer MBT flakes using circularly polarized light. This special property can seed the development of topological spintronics based on edge-state chirality switching (*73*, *74*) using ultrafast optical excitations.

## Materials and Methods

### Sample growth

The MBT single crystals were grown using a self-flux method (75). The mixtures of high-purity manganese powder (99.95%), bismuth shot (99.999%), and tellurium ingot (99.9999+%) with the molar ratio of Mn:Bi:Te = 1:10:16 are heated up to 900 °C for 12 hours to promote homogeneous melting and slowly cooled down (1.5 °C/h) to a temperature within the 590 °C–630 °C range, followed by centrifugation to remove excess flux. The MBT single crystals were cleaved *in situ* under a pressure $< 5 \times 10^{-11}$ mbar for the ARPES measurements.

The thin-film materials $(MnBi_2Te_4)/(Bi_2Te_3)_{30}$ and $(Bi_2Te_3)_{27}$ (Supplementary Text 3) were grown by molecular beam epitaxy using 99.9998% Mn, 99.999% Bi, and 99.9999% Te. $Bi_2Te_3$ films were grown on 0.05 wt% Nb-doped $SrTiO_3$ (111) substrates at 240 °C with a Bi:Te flux ratio of 1:16. The $MnBi_2Te_4$ top layer was formed by depositing MnTe on top of $Bi_2Te_3$ and annealed at 270 °C in a Te-rich atmosphere. The films were then transferred *in situ* to the ARPES chamber for measurements.

### ARPES measurements

The static and time-resolved (tr)-ARPES measurements were performed on the Multi-Resolution Photoemission Spectroscopy (MRPES) platform at the University of Chicago (76). The 6 eV laser for static ARPES was generated from a mode-locked Ti:sapphire oscillator with a repetition rate of 80 MHz. The trARPES setup featured a 200 kHz Yb:KGW laser accompanied by non-collinear optical parametric amplifiers to produce ultrafast 1.5 eV IR pump pulses and 6 eV ultraviolet probe pulses. The energy resolutions of the static and time-resolved ARPES setups were better than 4 meV and 20 meV, respectively. Focused probe beam waists, as characterized by the full-widths-half-maximum, were $14 \times 20$ μm$^2$ and $34 \times 53$ μm$^2$ for the static and time-resolved ARPES experiments, respectively. A systematic alignment procedure was adopted to ensure the overlap of the probed regions for static and time-resolved ARPES (22). The linearly polarized, $\sim 110 \times 140$ μm$^2$-sized, ~20 fs-long IR pump pulses were dimmed to incident fluences of 20 μJ/cm$^2$ and below. The time resolution was determined to be ~150 fs, limited by the duration of the probe pulses.

### trMOKE measurements

Steady state RMCD measurements were done with 1.95 eV continuous wave (CW) laser and with the pulsed 1.55 eV Ti:sapphire output (Coherent Chameleon) respectively. The light beam was modulated at 50 kHz between the left and right circular polarization using a photoelastic modulator (Hinds PEM). The reflected light was focused onto a photodiode. The MCD was determined as the ratio of the ac component of the photodiode signal measured by a lock-in amplifier at the polarization modulation frequency and the dc component of the photodiode signal measured by a voltmeter. In trMOKE, the probe beam was the output of the Ti:sapphire laser at 1.55 eV, and the pump beam was the second harmonic of an optical parametric oscillator (Coherent Compact OPO) at 1.88 eV. The time delay between the pump and probe pulses was controlled by a motorized linear delay stage, and the pump was modulated with a mechanical chopper. The reflected probe light passed through a half-wave Fresnel rhomb and a Wollaston prism and detected by a balanced photodiodes locked at the chopper frequency. The pump light spot diameter was around 1.5 μm. The optical measurements were done on a ~100-nm-thick sample flake on a $SiO_2$/Si substrate in a microscopic optical cryostat (attoDry 1000) with a base temperature of 3.5 K and a superconducting solenoid magnet up to 9 T.

## Acknowledgments

We thank Jinliang Ning from Tulane University for providing details of the published data on the heat capacity calculation. We also thank Yunhe Bai, Jiwoong Park, and Peter Littlewood at the University of Chicago, as well as Heike Pfau and Cuizu Chang at Pennsylvania State University for helpful discussions.

## Funding:

This work was supported by the U.S. Department of Energy DE-SC0022960 (KDN, WL, GB, CY, CII, HL, QG, SY), the U.S. Department of Energy DE-SC0022983 (JD, TW, XXZ), and National Science Foundation through the Penn State 2D Crystal Consortium-Materials Innovation Platform 2DCC-MIP under NSF Cooperative Agreement DMR 2039351 (SHL, ZM).

## Author contributions:

Conceptualization and supervision: SY

μARPES and trARPES measurements: KDN, WL, GB, CY, CII, HL, QG, SY

MBE growth: KDN, WL

trMOKE experiments: JD, TW, XXZ

Single crystal growth: SHL, ZM

Theoretical support: CL, BY

Data analysis and theoretical calculations: KDN

Writing—original draft: KDN, SY

Writing—review & editing: KDN, SY, XXZ, CL, ZM, GB, SHL, JD, BY

## Competing interests:

Authors declare that they have no competing interests.

## Data and materials availability:

All data needed to evaluate the conclusions in the paper are available in the main text or the supplementary materials.

# Figures and Tables

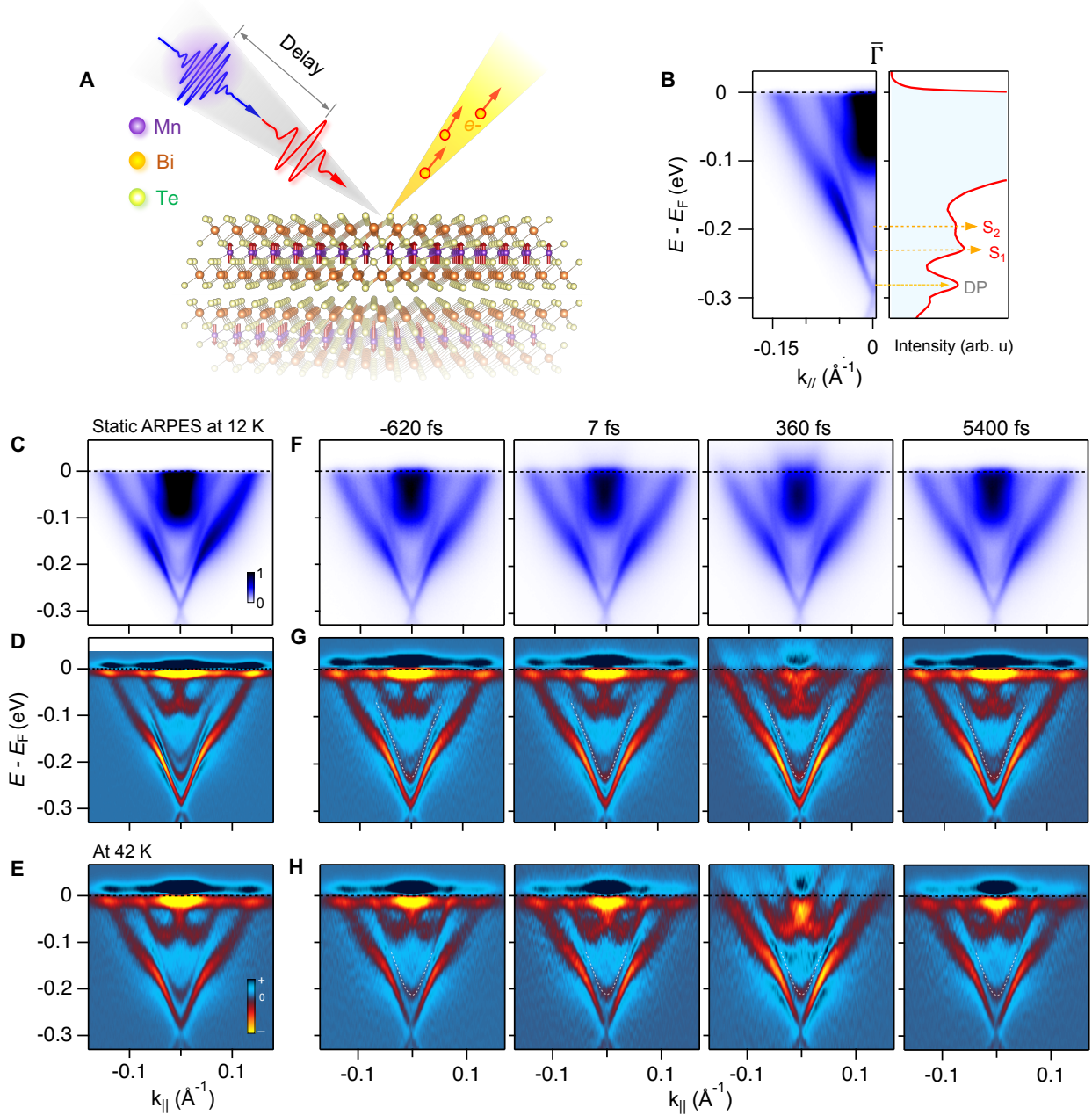

**Fig. 1. Evolution of the electronic structure in MnBi$_2$Te$_4$ (MBT) resolved by trARPES.** (**A**) Scheme of trARPES experiment on MBT. (**B**) The static ARPES spectrum of MBT along the $\overline{\Gamma} - \overline{K}$ direction (left) and the energy distribution curve (EDC) taken at $\overline{\Gamma}$ (right) with quasi-2D sub-bands S$_1$, S$_2$, and Dirac point marked. (**C**) Full ARPES spectrum at 12 K. Second derivative plots at (**D**) 12 K and (**E**) 42 K are also shown. (**F**) Time-dependent ARPES spectra at select delays and the base temperature of 12 K for an incident pump fluence of 10 μJ/cm$^2$. Time-dependent, second-derivative spectra at (**G**) 12 K and (**H**) 42 K are also shown.

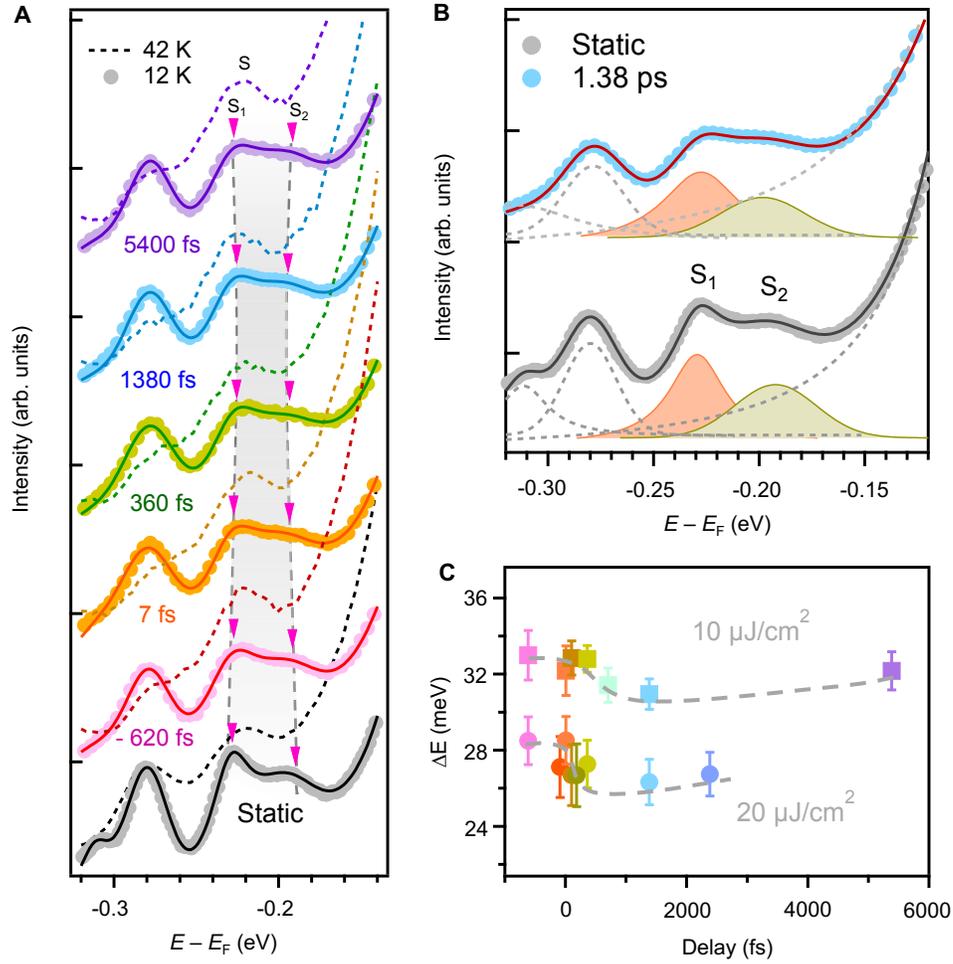

**Fig. 2. Analysis of energy distribution curves (EDCs) at the $\bar{\Gamma}$ point.** (**A**) EDCs taken at $\bar{\Gamma}$ for select delays, with the pump fluence of 10 µJ/cm². Results for the base temperature of 12 K (solid balls) and 42 K (dashed lines) are directly compared. Solid curves denote the fit curves using a five-Lorentzian model to extract the exchange gap (pink triangles) at each delay. (**B**) Exemplary fits to the EDC taken with static ARPES (grey balls), and that with trARPES at the delay of 1.38 ps (blue balls). (**C**) Time-dependent exchange gaps between the $S_1$ and $S_2$ sub-bands for the IR fluences of 10 and 20 µJ/cm². The dashed lines are guides to the eyes. Error bars show one standard deviation of uncertainty of the fitting.

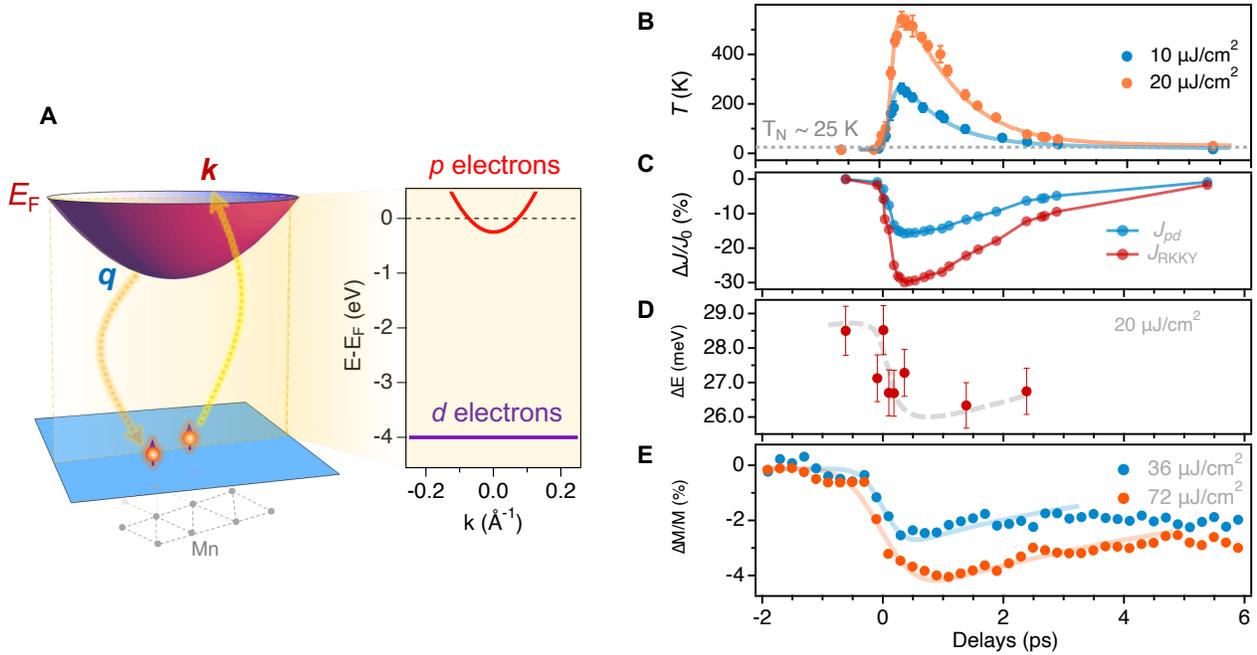

**Fig. 3. Magnetic interactions in a 2D system with itinerant *p* electrons and localized Mn *d* electrons**. (**A**) Illustration of an RKKY model which assumes the indirect interaction between localized *d* electrons via itinerant *p* electrons in two dimensions. (**B**) Transient electronic temperatures extracted by fitting the momentum-integrated energy distribution curve (EDC) at each delay to a modified Fermi-Dirac function (Supplementary Text 1). Error bars show standard deviation of the fitting. Solid curves denote simulation results using a microscopic Boltzmann model (Supplementary Text 2). The resulting *p-d* and RKKY coupling strengths are calculated and summarized in (**C**) based on the transient electronic temperature using a pump fluence of 20 μJ/cm$^2$. (**D**) Time-dependent exchange gap in the *q*-2DS also using the pump fluence of 20 μJ/cm$^2$. (**E**) Photo-induced demagnetization measured by trMOKE using a pump fluence of 36 and 72 μJ/cm$^2$. Solid lines indicate the fitting curves using an exponential decay convolved with a Gaussian function.

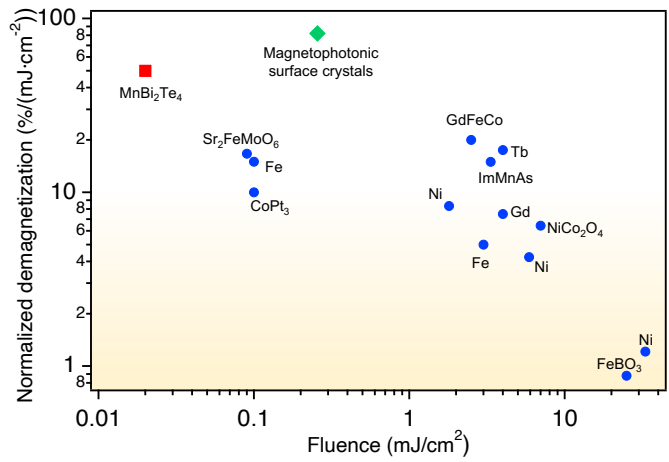

**Fig. 4. Summary of the percentages of demagnetization normalized by incident pump fluences, measured by trMOKE on MBT and some common magnets** *(60–71)*.